\documentclass[conference]{IEEEtran}
\IEEEoverridecommandlockouts
\def\FIGDIR{./Figures}          % directory of figures

\usepackage{cite}
\usepackage{amsmath,amssymb,amsfonts}
\usepackage{algorithmic}
\usepackage{graphicx}
\usepackage{textcomp}
\usepackage{xcolor}
\usepackage{multirow}
\usepackage{hyperref}

\definecolor{darkgreen}{RGB}{0,120,0}

\newcommand{\insertFigure}[3]{
    \begin{figure}[t]
        \centering
        \includegraphics[width=\linewidth]{\FIGDIR/#1.pdf}
        \caption{\small #2}
        \label{fig:#1}
    \vspace{#3}        
    \end{figure}
}

\newcommand{\betterparagraph}[1]{\noindent \textbf{#1. }}

\title{Inter-Layer Scheduling Space Exploration for Multi-model Inference on Heterogeneous Chiplets}

\author{
\IEEEauthorblockN{
Mohanad Odema,
Hyoukjun Kwon,
Mohammad Abdullah Al Faruque}

Department of Electrical Engineering and Computer Science \\
University of California, Irvine, USA \\
{\{modema, hyoukjun.kwon, alfaruqu\}}@uci.edu
\vspace{-5truemm}
}

\def\BibTeX{{\rm B\kern-.05em{\sc i\kern-.025em b}\kern-.08em
    T\kern-.1667em\lower.7ex\hbox{E}\kern-.125emX}}

\begin{document}

\maketitle

\begin{abstract}

To address increasing compute demand from recent multi-model workloads with heavy models like large language models, we propose to deploy heterogeneous chiplet-based multi-chip module (MCM)-based accelerators. We develop an advanced scheduling framework for heterogeneous MCM accelerators that comprehensively consider complex heterogeneity and inter-chiplet pipelining. Our experiments using our framework on GPT-2 and ResNet-50 models on a 4-chiplet system have shown upto 2.2$\times$ and 1.9$\times$ increase in throughput and energy efficiency, compared to a monolithic accelerator with an optmized output-stationary dataflow.

%This work studies and analyzes how exploring the inter-layer scheduling space on heterogeneous multi-chiplet modules (MCMs) can enhance performance efficiency for multi-model inference. 
% Prior works have studied the merit of cross-layer pipelining on homogeneous MCMs.  
%Our key contribution lies in characterizing the inter-layer scheduling space to exploit the synergy between chiplets' dataflow heterogeneity and inter-layer pipelining, which lays the foundation for a more generic scheduling optimization framework that can scale with the size and diversity of both the multi-models and the MCM architecture. Our experiments on GPT-2 and ResNet-50 models on a 4-chiplet system have shown that up to 2.2$\times$ increase in throughput and 1.9$\%$ increase in energy efficiency can be realized through leveraging heterogeneity and pipelining. 

\end{abstract}

\begin{IEEEkeywords}
heterogeneous chiplets, multi-chiplet module, multi-model, pipelining
\end{IEEEkeywords}
\section{Introduction}
\label{sec:introduction}

Recent artificial intelligence (AI) inference workloads have increased their scale in both of the model size (e.g., large language models~\cite{brown2020language, touvron2023llama}) and the number of models deployed together (e.g., AR/VR~\cite{kwon2023xrbench}), which constructs multi-model workloads with heavier models than those in the past. Such trends have led to heavier demands on compute capabilities in AI hardware from edge to cloud devices. As an approach to increase the compute capability in each chip for AI, chiplet-based multi-chip module (MCM) package has emerged as a promising solution~\cite{shao2019simba, tan2021nn, wang2022ai, orenes2023massive}. Such MCM packages facilitate the scaling of chips based on their composability, unlike monolithic designs whose scalability is often constrained by fabrication yields, power, and heat, and other engineering costs such as verification.

Researchers have actively explored the MCM design space for AI, focusing on the compiler mapping (or, dataflow) and workload orchestration onto MCMs considering the network-on-package (NoP) and other communication constraints~\cite{shao2019simba, tan2021nn, wang2022ai, orenes2023massive}. For example, Simba~\cite{shao2019simba} proposed a scalable MCM inference architecture that enables chiplets to either act as standalone inference engines or collaborate as groups for each layer in a neural network. Although Simba and other previous works have successfully delivered superior performance and energy efficiency than monolithic designs by thoroughly considering the underlying MCM architecture, they focus on single-model workloads targeting \textit{homogeneous} chiplets, while multi-model workloads with diverse models are emerging.

Therefore, considering the new AI workload trend, we propose to explore heterogeneous chiplet-based MCM packages as a solution, where each chiplet has an AI accelerator with different dataflow to address the workload heterogeneity in multi-model workloads. In addition to the heterogeneity, we consider inter-layer pipelining to reduce the amount offchip traffic and extensively explore the resulting scheduling space, which consists of heterogeneity-aware chiplet assignment for each layer and inter-chiplet pipelining, as illustrated in~\autoref{fig:Overview}. We implement a MAESTRO~\cite{kwon2019understanding}-based cost model targeting heterogeneous MCM to enable fast scheduling space exploration.

Using our scheduling framework, we perform a preliminary evaluation using a 2x2 heterogeneous chiplet MCM targeting a multi-model workload running GPT-2~\cite{radford2019language} large language model (LLM) and ResNet-50~\cite{he2015deep} image classification model together. In our evaluation , the heterogeneity and our scheduler together improves the throughput and the energy efficiency by up to 2.2 $\times$ and 1.9 $\times$, respectively, compared to a monolithic accelerator running an output-stationary (OS) dataflow. The results indicate that the heterogeneous MCM can be an effective approach for emerging multi-model workload with heavy models like LLMs.

In this abstract, we illustrate our framework at a high level and discuss preliminary evaluation results to provide more insights into our preliminary results and the potential of our proposed approach.

\insertFigure{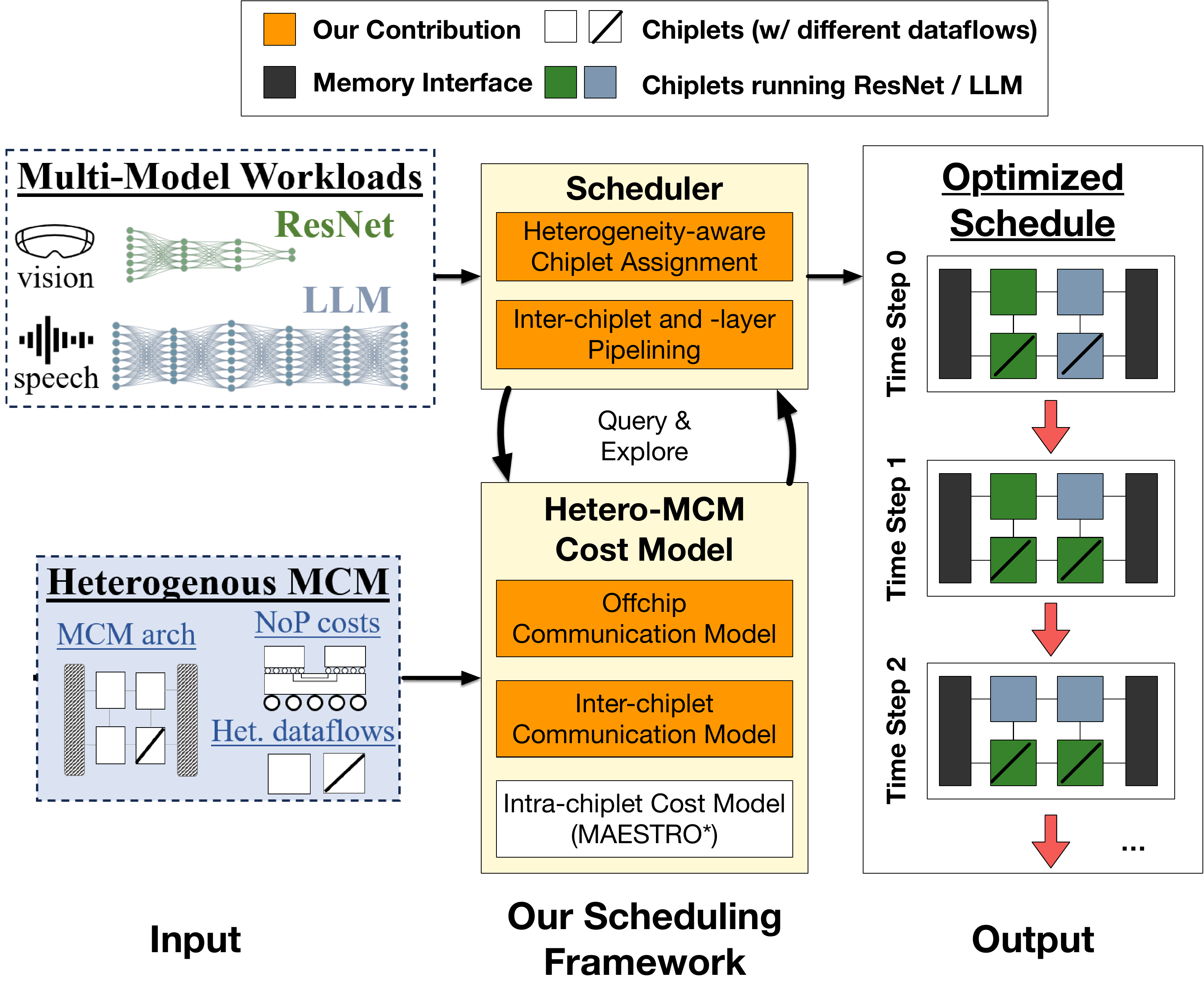}{Illustration of the generic scheduling optimization framework targeting multi-model workloads operation on heterogeneous MCMs}{-4ex}

\section{Scheduling Framework}
\label{sec:framework}

\betterparagraph{Cost Modeling} We model heterogeneous MCM architectures that have memory interfaces at the left and right sides, as illustrated in the ouptut of ~\autoref{fig:Overview}. Each chiplet on the left- and right-most columns possess a direct link to the offchip DRAM via double sided memory channels. We construct a cost model considering the NoP and offchip communication overheads utilizing the architectural parameters from \cite{shao2019simba, orenes2023massive} as shown in Table \ref{tab:params}. We deploy heterogeneous dataflows in the chiplets to support diverse workloads in multi-model workloads. In our preliminary study, we deploy output- and weight-stationary (os and ws) dataflows and utilize an open-source accelerator cost model, MAESTRO\cite{kwon2019understanding, kwon2020maestro}, for evaluating intra-chiplet performance based on the dataflow choices. By default, we set the size of the global buffer in each chiplet to be 10 MB, inspired by the on-chip memory size in a recent mobile accelerator~\cite{qualcomm_hexagon_680}, which can be modified by users.

\betterparagraph{Scheduling} We implement a two-stage scheduling framework. The first stage assigns desired chiplets for each layer considering their favored dataflow. The second stage explores the complex scheduling space of inter-layer pipelining, where we utilize the RA-tree~\cite{cai2023inter} structure to represent complex inter-layer scheduling space, apply a heuristic-driven (e.g., place starting node to be one adjacent to a memory interface channel) optimization algorithm to narrow-down schedule candidates, and select the best schedule among those candidates.

%specify heterogeneity in the accelerator chiplets through the choice of dataflow configuration can provide varying performance trade-offs based on the workload computational properties. In our case, we consider \textbf{\textit{output-stationary (os)}} and \textbf{\textit{weight-stationary (ws)}} dataflows, and use the open-source tool MAESTRO \cite{kwon2020maestro} for our intra-chiplet performance evaluation based on the chosen dataflow, on-chiplet resources, and model properties (e.g., layer shapes and sizes). Each chiplet's global buffer size is set to 10 MB.

% characterize heterogeneity on the accelerator chiplets via the classes of data, namely \textbf{\textit{output-stationary (os)}} and \textbf{\textit{weight-stationary (ws)}} accelerators. The naming terms are used as such to reflect the class of \textit{dataflow} configuration supported by the accelerator resource, where based on the properties of the scheduled workload (e.g., the dataflow affinities, computational intensity, and parallelization potential), different performance trade-offs can be achieved. 

\begin{table}[t]
    \centering
    \caption{MCM microarchitecture parameters from \cite{shao2019simba, orenes2023massive}. All numbers are scaled to 28 nm process node technology.}
    \vspace{-2ex}
    \begin{tabular}{l | c | c}
    \hline
    \multirow{3}{*}{Package} & NoP interconnect latency & 35 ns/hop \\
    & NoP interconnect energy & 2.04 pJ/bit \\
    & NoP interconnect bandwidth & 100 GB/s/Chiplet \\
    \hline
    \multirow{3}{*}{Offchip Memory} & DRAM latency & 200 ns \\
    & DRAM energy & 14.8 pJ/bit \\
    & DRAM bandwidth & 64 GB/s \\
    \hline
    % \multirow{3}{*}{Chiplet} & Number of processing engines & 256 \\
    % & MAC energy & 0.00023 nJ \\
    % & NoC latency & \\
    % & NoC ergy & \\
    % & NoC bandwidth &  \\ \hline
    
    \end{tabular}
    \label{tab:params}
    \vspace{-3ex}
\end{table}

\section{Evaluation}
\label{sec:evaluation}

\begin{figure}[!tbp]
\begin{center}
{\includegraphics[,width = 0.48\textwidth]{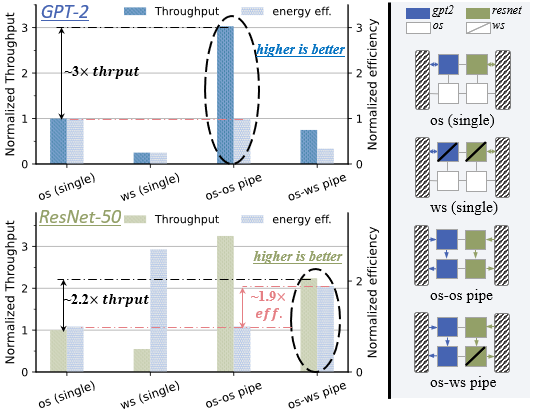}}
\end{center}
\vspace{-2.6ex}
\caption{\textbf{\textit{Left:}} Throughput and energy efficiency scores for the various scheduling options of the GPT-2 and ResNet-50 normalized against the scores from the standalone \textbf{os} option. \textbf{\textit{Right:}} Illustration of the considered schedules.}
\label{fig:exp}
\vspace{-3ex}
\end{figure}

\betterparagraph{Heterogeneous MCM Configuration} We simulate a scenario where MCM chiplets operating at 500 MHz are arranged in 2$\times$2 mesh topology in an NoP fashion. Based on the architecture, we evalute 4 candidates for inter-layer scheduling space: \textbf{os}, \textbf{ws}, \textbf{os-os}, and \textbf{os-ws}. The first options, \textbf{os} and \textbf{ws}, reflect the \textit{standalone} scheduling options on a single chiplet.
% , leveraging intra-layer parallelism opportunities conditioned on the available resource budget and the model computational properties.
% while incurring the necessary data swapping and memory read/write based on the chiplet's resource constraints and the size of data. 
The other two are options that additionally leverage inter-layer pipelining through utilizing multiple chiplets to process different stages of a model, which can lead to improved resource utilization in the cases when workloads lack sufficient degrees of parallelism within. The \textbf{os-os} option represents an instance of \textit{homogeneous} pipelining in which same type accelerators are employed for the various pipeline stages, as in SIMBA \cite{shao2019simba}. The \textbf{os-ws} is an instance of heterogeneous chiplets' pipelining. Figure \ref{fig:exp} illustrates this inter-layer scheduling space.

\betterparagraph{Workload} We employ an LLM, GPT-2~\cite{radford2019language}, and a computer vision model, ResNet-50~\cite{he2015deep}, as a multi-model workload example. Due to the sheer difference in their computing overheads, we perform our analysis using a single layer of the GPT-2 model as per their definition of layer, which constitutes a number of computing sublayer blocks within \cite{vaswani2017attention}. We construct 2-stage pipelining scheduling options by partitioning our models at layers that  provide comparable energy-delay product (EDP) and latency for both stages. We assume the batch size of 1.

\betterparagraph{Metric} We employ \textit{throughput} and \textit{efficiency} as our evaluation metrics. We measure the \textit{throughput} as the number of computed outputs for each pipeline latency (i.e., $\frac{\# outputs}{sec}$). We measure the \textit{efficiency} as the inverse of the EDP.

%outcome from the product of number of batches and number of pipeline stages divided by the latency of the slowest pipeline stage. Whereas \textit{efficiency} is defined as the inverse of the EDP, where EDP is computed as the sum of energy-latency products at each pipeline stage. The inter-chiplet and offchip transmission costs are also considered as separate pipelining stages in our evaluation. For both metrics, \textit{higher values are better}.

\betterparagraph{Results} We summarize the evaluation results in~\autoref{fig:exp}. Because GPT-2 model repeats the same building blocks (e.g., the matrix size in attention blocks is the same across the model), the heterogeneity is not helpful if we just run the GPT-2. The results suggest that output stationary is friendly to the building blocks. However, the pipelining enabled significant improvements in the throughput up to 3$\times$ compared to the monolithic os accelerator. 

We observe similar benefits of pipelining in ResNet-50, which increased the throughput by 3.1$\times$ by applying pipelining. However, we also observe that the heterogeneity can provide significant efficiency benefits (1.9 $\times$ higher than the os accelerator) at the cost of some throughput compared to the pipelined homogeneous chiplet. Such results imply a new trade-off space between the efficiency and throughput enabled by the heterogeneity. Overall, the heterogeneity and pipelining via our scheduler together improve the throughput and efficiency by up to 2.2 $\times$ and 1.9 $\times$, respectively.

% \HK{Please clarify what "1" means (i.e., normalized against what?)} \HK{Thanks for the clarificaiton. Please explicitly say within the caption or the figure that those data are normalized against the standalone os} 

% \blue{both are normalized with respect to the standalone os option scores so unitless? I mentioned the normalization at the start of the paragraph starting with `Figure 2..`. without normalization throughput is batches/second and efficiency is $(kJ.s)^{-1}$}

\section{Conclusion}
\label{sec:conclusion}

We have explored the synergy of chiplet heterogeneity and advanced scheduling with pipelining and shown the efficacy of the approach using a preliminary evaluation. The results indicate a new trade-off space between the performance and energy based on the heterogeneity and pipelining with promising results. As future works, we will explore more extensive search space with various MCM hardware scales and more workloads to show the effectiveness of our proposed approach.

%We have shown how the synergy of chiplet heterogeneity and pipelining parallelism can enhance performance efficiency on shared MCM resource servicing multi-models. Still, the optimization space complexity for scheduling multi-models on heterogeneous MCMs rises significantly when the problem scales on the architecture, workloads and heterogeneity levels. Thus, the next step along this research direction is to implement a generic, scalable scheduling optimization framework.

% It should be noted that in general, other forms of chiplet heterogeneity exist via varying levels of computational capabilities and functionalities.

\bibliographystyle{ieeetr}
\bibliography{ref}

\end{document}